\documentclass{article}

\usepackage{arxiv}

\usepackage[utf8]{inputenc} 
\usepackage[T1]{fontenc}    
\usepackage{hyperref}       
\usepackage{xurl}           
\usepackage{booktabs}       
\usepackage{amsfonts}       
\usepackage{nicefrac}       
\usepackage{microtype}      
\usepackage{graphicx}
\usepackage[square,numbers]{natbib}
\usepackage{doi}

\title{Comparing the Utility and Disclosure Risk of Synthetic Data with Samples of Microdata}

\author{{Claire Little}\\
	School of Social Sciences\\
	University of Manchester\\
	Manchester, M13 9PL, UK \\
	\texttt{claire.little@manchester.ac.uk} \\
	\And
	{Mark Elliot} \\
	School of Social Sciences\\
	University of Manchester\\
	Manchester, M13 9PL, UK \\
	\texttt{mark.elliot@manchester.ac.uk} \\
	\And
	{Richard Allmendinger} \\
	Alliance Manchester Business School\\
	University of Manchester\\
	Manchester, M13 9PL, UK \\
	\texttt{richard.allmendinger@manchester.ac.uk} \\
}

\date{}

\renewcommand{\headeright}{Preprint}
\renewcommand{\shorttitle}{Comparing the Utility and Disclosure Risk of Synthetic Data with Samples of Microdata}

\begin{document}
\maketitle

\begin{abstract}
	Most statistical agencies release randomly selected samples of Census microdata, usually with sample fractions under 10\% and with other forms of statistical disclosure control (SDC) applied. An alternative to SDC is data synthesis, which has been attracting growing interest, yet there is no clear consensus on how to measure the associated utility and disclosure risk of the data. The ability to produce synthetic Census microdata, where the utility and associated risks are clearly understood, could mean that more timely and wider-ranging access to microdata would be possible.
	
	This paper follows on from previous work by the authors which mapped synthetic Census data on a risk-utility (R-U) map. The paper presents a framework to measure the utility and disclosure risk of synthetic data by comparing it to samples of the original data of varying sample fractions, thereby identifying the sample fraction which has equivalent utility and risk to the synthetic data. Three commonly used data synthesis packages are compared with some interesting results. Further work is needed in several directions but the methodology looks very promising.
\end{abstract}

\keywords{Data Synthesis \and Microdata \and Disclosure Risk \and Data Utility}

\section{Introduction}
Many statistical agencies release randomly selected Census samples to researchers (and sometimes publicly), usually with sample fractions under 10\% and with other forms of statistical disclosure control (SDC)~\cite{hundepool_statistical_2012} applied. However, as noted by Drechsler et al.~\cite{Drechsler2010SamplingMicrodata}, intruders (or malicious users) are becoming more sophisticated and agencies using standard SDC techniques may need to apply them with higher intensity, leading to the released data being of reduced quality for statistical analysis. 

An alternative to SDC is data synthesis \cite{Rubin1993StatisticalLimitation,Little1993StatisticalData}, which takes original data and produces an artificial dataset with the same structure and statistical properties as the original, but that (in the case of fully synthetic data) does not contain any of the original records. As the data is synthetic, attributes which are normally suppressed, aggregated or top-coded (such as geographical area or income) may then be included, allowing more complete analysis. As no synthetic record should correspond to a real individual, fully synthetic data should present very low disclosure risk -- the risk of re-identification is not meaningful, although there is still likely to be a residual risk of attribution~\cite{Taub2018DifferentialExploration}. Interest in synthetic data is growing, yet there is no clear consensus on how to measure the associated utility and disclosure risk of the data, such that users may have an understanding of how closely a synthetic dataset relates to the original data. 

This paper follows on from previous work~\cite{Little2021GenerativeAdversarial}, which mapped synthetic Census data on the risk-utility (R-U) map, and presents a framework to measure the utility and disclosure risk of synthetic data by comparing it against random samples of the original data of varying sample fractions, thereby identifying the sample fraction equivalence of the synthetic dataset. For instance, a particular synthetic dataset might be equivalent in terms of utility to a 20\% sample of the original data, and in terms of disclosure risk to a 10\% sample. Since Census microdata tends to be released with sample fractions between 1\% to 10\%, the ability to determine how a synthetic dataset compares in terms of utility and disclosure risk would allow data producers a greater understanding of the appropriateness of their synthetic data.  
To test our framework, we performed experiments using four Census microdata sets with synthetic data generated using three state-of-the-art data synthesis methods (\textit{CTGAN}~\cite{Xu2019ModelingGAN}, \textit{Synthpop}~\cite{Nowok2016Synthpop:R} and \textit{DataSynthesizer}~\cite{Ping2017DataSynthesizer:Datasets}). 

Section \ref{Background} provides a brief introduction to the data synthesis problem, particularly in terms of microdata, and an introduction to the data synthesis methods. Section \ref{Design} outlines the design of the study, describing the methods and the Census data used. Section \ref{Results} provides the results whilst Section \ref{Discussion} considers the findings, and Section \ref{Conclusion} concludes with directions for future research.

\section{Background}
\label{Background}
\subsection{Data Synthesis}

Rubin~\cite{Rubin1993StatisticalLimitation} first introduced the idea of synthetic data as a confidentiality protection mechanism in 1993, proposing using multiple imputation on all variables such that none of the original data was released. In the same year, Little~\cite{Little1993StatisticalData} proposed an alternative that simulated only sensitive variables, thereby producing partially synthetic data. Rubin's idea was slow to be adopted, as noted by Raghunathan et al.~\cite{Raghunathan2003MultipleLimitation}, who along with Reiter~\cite{Reiter2002SatisfyingSets,Reiter2003InferenceSets,Reiter2003ReleasingStudy}, formalised the synthetic data problem. Further work (e.g.~\cite{Reiter2005UsingMicrodata,Drechsler2010SamplingMicrodata,Drechsler2011AnDatasets}) has involved using non-parametric data synthesis methods such as classification and regression trees (CART) and random forests and more recently deep learning methods such as Generative Adversarial Networks~\cite{Goodfellow2014GenerativeNets} have also been used to generate synthetic data.

There are two competing objectives when producing synthetic data: high data utility (i.e., ensuring that the synthetic data is useful, with a distribution close to the original) and low disclosure risk. Balancing this trade-off can be difficult, as, in general, reducing disclosure risk comes with a concomitant cost for utility. This trade-off can be visualised using the R-U confidentiality map developed by Duncan et al.~\cite{Duncan2004DatabaseMap}. There are multiple measures of utility, ranging from comparing summary statistics, correlations and cross-tabulations, to considering data performance using predictive algorithms. However, for synthetic data there are fewer that measure disclosure risk. As noted by Taub et al.~\cite{Taub2018DifferentialExploration}, much of the SDC literature focuses on re-identification risk, which is not meaningful for synthetic data, rather than the attribution risk, which is relevant. The Targeted Correct Attribution Probability (TCAP)~\cite{Elliot2014FinalTeam,Taub2018DifferentialExploration} can be used to assess attribution risk.

\subsection{Synthetic Census Microdata}
\label{Synthetic}
Since Census microdata is predominantly categorical, it requires synthesis methods that can handle categorical data. CART, a non-parametric method developed by Breiman et al.~\cite{Breiman1984ClassificationTrees}, can handle mixed type (and missing) data, and can capture complex interactions and non-linear relationships. CART is a predictive technique that recursively partitions the predictor space, using binary splits, into relatively homogeneous groups; the splits can be represented visually as a tree structure, meaning that models can be intuitively understood (where the tree is not too complex). Reiter~\cite{Reiter2005UsingMicrodata} used CART to generate partially synthetic microdata, as did Drechsler and Reiter~\cite{Drechsler2010SamplingMicrodata}, who replaced sensitive variables in the data with multiple imputations and then sampled from the multiply-imputed populations. Random forests, developed by Breiman~\cite{breiman2001random}, is an ensemble learning method and an extension to CART in that the method grows multiple trees. Random forests were used to synthesise a sample of the Ugandan Census~\cite{Drechsler2011AnDatasets} and to generate partially synthetic microdata~\cite{caiola2010random}.

\textit{Synthpop}, an open source package written in the R programming language, developed by Nowok et al.~\cite{Nowok2016Synthpop:R}, uses CART as the default method of synthesis (other options include random forests and various parametric alternatives). Since CART is a commercial product, \textit{Synthpop} uses an open source implementation of the algorithm provided by the \textit{rpart} package~\cite{rpart2019}. \textit{Synthpop} synthesises the data sequentially, one variable at a time; the first is sampled, then the following are synthesised using the previous variables as predictors. Whilst an advantage of \textit{Synthpop} is that it requires little tuning and performs very quickly, a disadvantage is that it (and tree-based methods in general) can struggle computationally with variables that contain many categories. As suggested by Raab et al.~\cite{Raab2017GuidelinesData} methods to deal with this include aggregation, sampling, changing the sequence order of the variables and excluding variables from being used as predictors. \textit{Synthpop} has been used to produce synthetic longitudinal microdata~\cite{Nowok2017ProvidingR} and synthetic Census microdata~\cite{Taub2020TheRecords,pistner2018synthetic}.

Another method that can process mixed type data is \textit{DataSynthesizer}, developed by Ping et al.~\cite{Ping2017DataSynthesizer:Datasets}, a Python package that implements a version of the PrivBayes~\cite{Zhang2017PrivBayes:Networks} algorithm. PrivBayes constructs a Bayesian network that models the correlations in the data, allowing approximation of the distribution using a set of low-dimensional marginals. Noise is injected into each marginal to ensure differential privacy and the noisy marginals and Bayesian network are then used to construct an approximation of the data distribution. PrivBayes then draws samples from this to generate a synthetic dataset. \textit{DataSynthesizer} allows the level of differential privacy to be set by the user, or turned off. It has been used to generate health data~\cite{Rankin2020ReliabilitySharing} and in exploratory studies~\cite{hittmeir2019utility,dankar2022multi-dimensional,nixon2022alatent}.

In the field of deep learning~\cite{lecun2015deeplearning}, Generative Adversarial Networks (GANs) have been generating much research interest and have been used for various applications, although as detailed by Wang et al.~\cite{Wang2020ANetworks} these are predominantly in the image domain. GANs, developed by Goodfellow et al.~\cite{Goodfellow2014GenerativeNets}, typically train two neural network (NN) models: a generative model that captures the data distribution and generates new data samples, and a discriminative model that aims to determine whether a sample is from the model distribution or the data distribution. The models are trained together in an adversarial zero-sum game, such that the generator goal is to produce data samples that fool the discriminator into believing they are real and the discriminator goal is to determine which samples are real and which are fake. Training is iterative with (ideally) both models improving over time to the point where the discriminator can no longer distinguish which data is real or fake. From a data synthesis perspective GANs are interesting in that the generative model does not access the original (or training) data at all, and starts off with only noise as input; in theory this might reduce disclosure risk. 

GANs for image generation tend to deal with numerical, homogeneous data; in general, they must be adapted to deal with Census microdata, which is likely to be heterogeneous, containing imbalanced categorical variables, and skewed or multimodal numerical distributions. Several studies have done this by adapting the GAN architecture, these are often referred to as tabular GANs (e.g.~\cite{Camino2018GeneratingNetworks,Park2018DataNetworks,Zhao2021CTAB-GAN:Synthesizing,Chen2019Faketables:Data}). \textit{CTGAN}, or Conditional Tabular GAN, developed by Xu et al.~\cite{Xu2019ModelingGAN} uses ``mode-specific normalisation'' to overcome non-Gaussian and multimodal distribution problems, and employs oversampling methods and a conditional generator to handle class imbalance in the categorical variables. In their study~\textit{CTGAN} outperformed Bayesian methods, and other GANs, for generating mixed type synthetic data.

National statistical agencies have released synthetic versions of microdata using forms of multiple imputation. The United States Census Bureau releases a synthetic version of the Longitudinal Business Database (SynLBD)~\cite{Kinney2011TowardsDatabase}, the Survey of Income and Program Participation (SIPP) Synthetic Beta~\cite{Benedetto2018TheBeta} and the OnTheMap application~\cite{Machanavajjhala2008Privacy:Map}. Whilst governmental organisations have not so far released synthetic microdata created using deep learning methods, research in this area is ongoing (e.g.~\cite{Kaloskampis2020SyntheticService,Joshi2019GenerativeClasses}).

\section{Research Design}
\label{Design}
To determine the sample equivalence, the risk and utility of generated synthetic Census data was compared to the risk and utility of samples of the original Census data (of various sample fractions).\footnote[1]{Note that, for calculating the risk and utility, the sample data was treated in the same way as the synthetic data, namely by comparing against the original data. However, for simplicity, in the metric descriptions only synthetic data is mentioned.} Four different Census microdata sets were used to demonstrate results on datasets from different underlying population data structures. Three state-of-the-art data synthesis methods were used to generate the synthetic data, each using the default parameters. To obtain consistent results, multiple datasets were generated (using different random seeds) for the sample and synthetic data, and the mean of the utility and risk metrics for each calculated. \footnote[2]{The project code is available here: \url{https://github.com/clairelittle/psd2022-comparing-utility-risk}}

\subsection{Data Synthesisers}
The methods used were \textit{Synthpop}~\cite{Nowok2016Synthpop:R}, \textit{DataSynthesizer}~\cite{Ping2017DataSynthesizer:Datasets} and \textit{CTGAN}~\cite{Xu2019ModelingGAN}. These were selected as they are established, open-source methods that should produce good quality data. Whilst the focus of these experiments was on evaluating the resulting utility and risk of the generated data, rather than the individual methods, for~\textit{DataSynthesizer} the differential privacy parameter was varied in order to understand how the use of differential privacy affects the quality and risk of the synthetic data and how such differentially private synthetic datasets compare to samples. For each parameter setting, five fully synthetic datasets were generated, each using a different random seed.
\\

\noindent \textbf{Synthpop} Version 1.7-0 of \textit{Synthpop} was used with default parameters. As described in Section \ref{Synthetic}, \textit{Synthpop} allows the sequence order of the variables to be set by the user, however there is no default for this (other than the ordering the data is in). Since the Census microdata used for these experiments is predominantly categorical, with many variables containing many categories, and it is known that \textit{Synthpop} can struggle with variables containing many categories, the visit sequence was set such that variables were ordered by the minimum to maximum number of categories, with numerical variables first (and a tie decided by ordering alphabetically). Moving variables with many categories to the end of the sequence is suggested by Raab et al.~\cite{Raab2017GuidelinesData}.
\\

\noindent \textbf{DataSynthesizer} Version 0.1.9 of \textit{DataSynthesizer} (described in Section~\ref{Synthetic}) was used with Correlated Attribute mode (which implements the PrivBayes~\cite{Zhang2017PrivBayes:Networks} algorithm). Default parameters were used, whilst differing the Differential Privacy (DP) parameter. DP is controlled by the $\epsilon$ parameter and a value of zero turns DP off. Four different values were used (DP = off, $\epsilon = 0.1$, 1, and 10). Lower values of $\epsilon$ tend to be used in practise, but the range of values aims to understand the effect at both the higher and lower end, as well as turning off DP altogether.
\\

\noindent \textbf{CTGAN} Version 0.4.3 of \textit{CTGAN} was used for all experiments. \textit{CTGAN} as described in Section~\ref{Synthetic}, is a Conditional GAN implemented in Python. There are many hyperparameters that might be altered for a GAN; the default values were used, with the number of epochs set at 300.

\subsection{Data}
\label{Data}
Four Census microdatasets were used, each from a different continent. Each dataset contains individual records, pertaining to adults and children. The variables include demographic information such as age, sex and marital status (i.e., variables that are often considered key identifiers) and a broad selection of variables pertaining to employment, education, ethnicity, family, etc. Each dataset contained the same key variables, and target variables that broadly cover the same overall themes. The purpose of using multiple datasets was not to directly compare the countries, but rather to determine whether any patterns uncovered during the experiments were replicated on similar (but not identical) datasets. Table~\ref{tab:data} describes the data in terms of sample size and features.

\begin{table}[t]
    \begin{centering}
    \caption{Census Data Summary}\label{tab:data}
    \begin{tabular}{lcccc}
        \toprule
        Dataset & Sample size & $\#$Total Variables & $\#$Categorical   & $\#$Numerical \\
        \midrule
        Canada 2011  & 32149     & 25    & 21    & 4 \\
        Fiji 2007    & 84323     & 19    & 18    & 1 \\
        Rwanda 2012  & 31455     & 21    & 20    & 1 \\
        UK 1991      & 104267    & 15    & 13    & 2 \\
        \bottomrule
    \end{tabular}
    
    \end{centering}
\end{table}

The data was minimally preprocessed and missing values were retained. Three of the datasets (UK~\cite{ONSCensusSARs}, Canada~\cite{IPUMsCensus} and Rwanda~\cite{IPUMsCensus}) were subsetted on a randomly selected geographical region; this was to reduce data size and also to naturally reduce the categories for some of the variables. The entire Fiji sample~\cite{IPUMsCensus} was used. Appendix A contains a summary of the datasets.
\\

\noindent \textbf{Creating the Census Data Samples} For each Census dataset, random samples (without replacement) of increasing sizes were drawn (0.1\%, 0.25\%, 0.5\%, 1\%, 2\%, 3\%, 4\%, 5\%, 10\%, 20\%, 30\%, 40\%, 50\%, 60\%, 70\%, 80\%, 90\%, 95\%, 96\%, 97\%, 98\%, 99\%). A focus was placed on those sample fractions closer to zero since released Census samples tend to be relatively small, and closer to 100\% to map the data as it became closer in size to the original. For each sample fraction 100 datasets were generated (with different random seeds), and the results of the individual risk and utility measures averaged. 

\subsection{Measuring Disclosure Risk Using TCAP}

Elliot~\cite{Elliot2014FinalTeam} and Taub et al.~\cite{Taub2018DifferentialExploration} introduced a measure for the disclosure risk of synthetic data called the Correct Attribution Probability (CAP) score. The disclosure risk is calculated using an adaptation used in Taub et al.~\cite{Taub2019challenge} called the \textit{Targeted Correct Attribution Probability} (TCAP). TCAP is based on a scenario whereby an intruder has partial knowledge about a particular individual. That is, they know the values for some of the variables in the dataset (the keys) and also that the individual is in the original dataset\footnote[3]{This is a strong assumption, which has the benefit of then dominating most other scenarios, the one possible exception is a presence detection attack. However, for Census data, presence detection is vacuous, and the response knowledge assumption is sound by definition.}, and wish to infer the value of a sensitive variable (the target) for that individual.  The TCAP metric is then the probability that those matched records yield a correct value for the target variable (i.e. that the adversary makes a correct attribution inference).

TCAP has a value between 0 and 1; a low value would indicate that the synthetic dataset carries little risk of disclosure whereas a TCAP score close to 1 indicates a higher risk. However, a baseline value can be calculated (this is essentially the probability of the intruder being correct if they drew randomly from the univariate distribution of the target variable) and therefore a TCAP value above the baseline might indicate some disclosure risk. Given this, we have chosen to scale the TCAP value between the baseline and a value of 1. This does create the possibility of a synthetic dataset receiving a negative TCAP score (which can still be plotted on the R-U map) but that simply indicates a risk level below that of the baseline. We refer to the scaled TCAP value as the marginal TCAP; i.e. it is the increase in risk above the baseline. 

For each Census dataset, three target and six key variables were identified and the corresponding TCAP scores calculated for sets of 3, 4, 5 and 6 keys. The overall mean of the TCAP scores was then calculated as the overall disclosure risk score. Where possible, the selected key/target variables were consistent across each country. Full details of the target and key variables are in Appendix B.

\subsection{Evaluating Utility}

Following previous work by Little et al.~\cite{Little2021GenerativeAdversarial} and Taub et al.~\cite{Taub2020TheRecords}, the utility of the synthetic and sample data was assessed using multiple measures. The confidence interval overlap (CIO) and ratio of counts/estimates (ROC) were calculated. This was to provide a more complete picture of the utility, rather than relying upon just one measure. The propensity score mean squared error (pMSE)~\cite{Snoke2018GeneralData} was not used as, whilst it is suitable for analysing the synthetic data it is not suited to the analysis of sample data as it is structurally tied to the original data (since the sample data is a subset of the original data).

To calculate the CIO (using 95\% confidence intervals), the coefficients from regression models built on the original and synthetic datasets are used. The CIO, proposed by Karr et al.~\cite{Karr2006FrameworkConf}, is defined as:
\begin{equation}{\textrm{CIO}} =  \frac{1}{2}\bigg\{\frac{\mbox{min}(u_{o},u_{s}) - \mbox{max}(l_{o},l_{s})}{u_{o}-l_{o}}+\frac{\mbox{min}(u_{o},u_{s}) - \mbox{max}(l_{o},l_{s})}{u_{s}-l_{s}}\bigg\}, \end{equation}
where $u_{o}$, $l_{o}$ and $u_{s}$, $l_{s}$ denote the respective upper and lower bounds of the confidence intervals for the original and synthetic data. This can be summarised by the average across all regression coefficients, with a higher CIO indicating greater utility (maximum value is 1 and a negative value indicating no overlap).
For each synthetic Census dataset, two logistic regressions were performed, with the CIO for each calculated. The mean of these two results (where a negative, or no overlap was counted as zero) was taken as the overall CIO utility score for that dataset. Details of the regression models for each dataset are presented in Appendix C.

Frequency tables and cross-tabulations are evaluated using the ROC, which is calculated by taking the ratio of the synthetic and original data estimates (where the smaller is divided by the larger one). Thus, given two corresponding estimates (for example, the number of records with sex = female in the original dataset, compared to the number in the synthetic dataset), where $y_{orig}$ is the estimate from the original data and  $y_{synth}$ is the corresponding estimate from the synthetic data, the ROC is calculated as:
\begin{equation} {\textrm{ROC}} = \frac{\mbox{min}(y_{orig}, y_{synth})}{\mbox{max}(y_{orig}, y_{synth})}.\end{equation}
If $y_{orig} = y_{synth}$ then the ROC = 1. Where the original and synthetic datasets are of different sizes (as is the case when calculating the ROC for the various sample datasets) the proportion, rather than the count can be used. The ROC was calculated over univariate and bivariate cross-tabulations of the data, and takes a value between 0 and 1. For each variable the ROC was averaged across categories to give an overall score. 

To create an overall utility score for comparing against the overall disclosure risk score (marginal TCAP), the mean of the ROC scores and the CIO was calculated -- a score closer to zero indicates lower utility; a score closer to 1 indicates higher utility.\footnote[4]{We recognise that averaging different utility metrics may not be optimal and in future work we will consider an explicitly multi-objective approach to utility optimisation.} The results for the synthetic and sample data were plotted on the R-U map for each country separately.

\section{Results}
\label{Results}
\begin{figure}[htb]
\includegraphics[width=\textwidth]{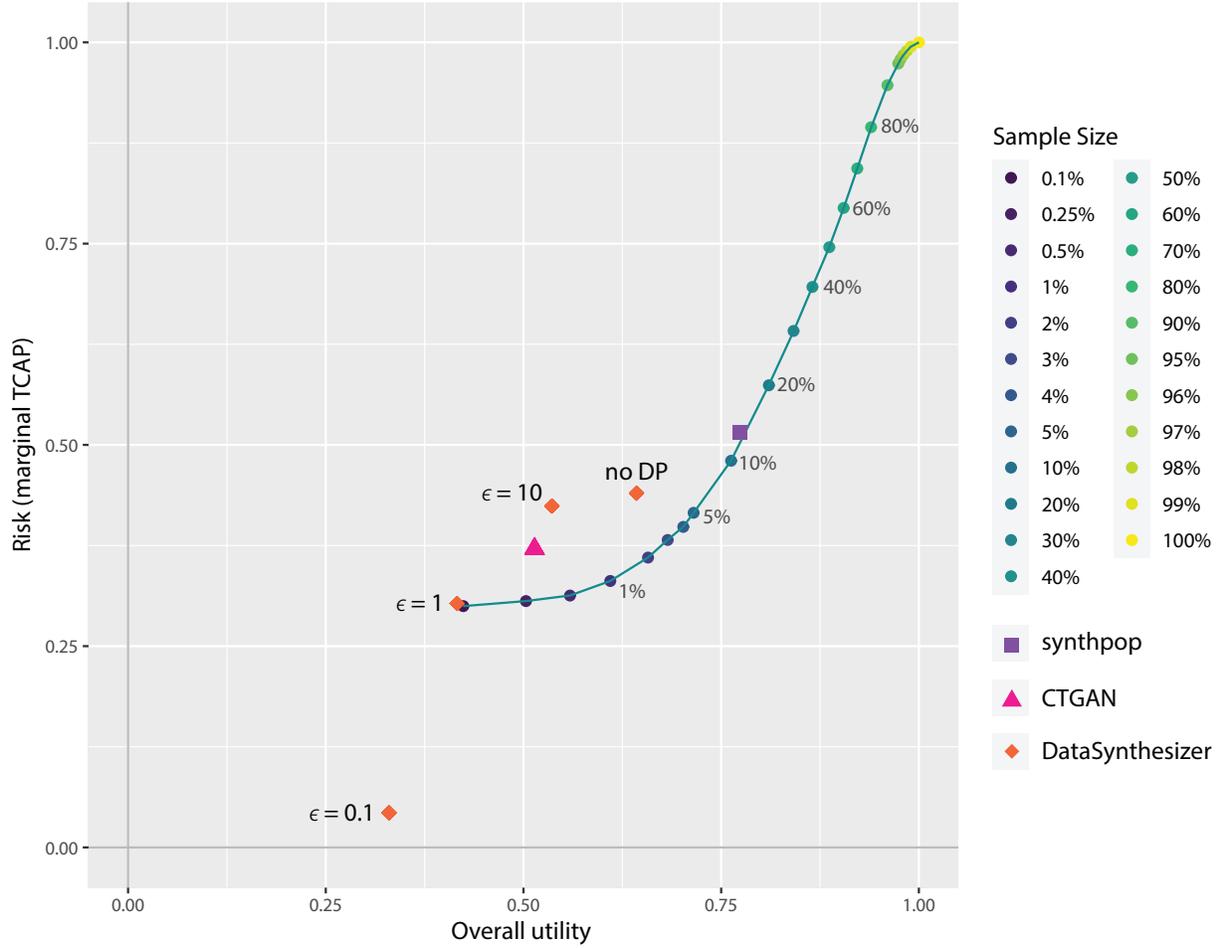}
\caption{Risk-Utility map plotting the mean synthetic data and sample fraction results for UK 1991 Census data. DP stands for Differential Privacy.} \label{uk1991ru}
\end{figure}

Fully synthetic datasets of the same size as the original were generated, and no post-processing was performed on the data. For \textit{Synthpop} and \textit{CTGAN} five different models were created (using different random seeds) and a synthetic dataset generated for each. The mean utility and risk over the five datasets is plotted on the R-U map. For \textit{DataSynthesizer} five different models for each value of $\epsilon$ were created and a point (the mean over the 5) is plotted for each of a series of values for $\epsilon$. Fig.~\ref{uk1991ru} illustrates the R-U map for the UK 1991 Census data. The sample fractions form a curved line, with a point representing each increasing sample fraction (from left to right). The utility starts to drop quite steeply once the sample fraction drops below about 3\%. Both the risk and utility of the data at 100\% (i.e. the whole original sample) is necessarily 1. The synthetic datasets are plotted alongside the different sample fractions to illustrate how they compare with the sample data. Considering Fig.~\ref{uk1991ru} the \textit{Synthpop} point falls almost on the sample curve, meaning it has utility and disclosure risk equivalence of between a 10\% and 20\% sample of the original data. Plots for the other three Census datasets are contained in Appendix D.

Table~\ref{tab:results} provides detail on the utility and risk (mean over five datasets) of the synthetic data for each of the Census datasets together with the associated sample fraction equivalence.\footnote[5]{Standard deviation not included for clarity as this was generally small, $<$0.01} A synthetic dataset with higher sample fraction equivalence for utility but lower sample fraction equivalence for risk would be optimal from the point of view of synthetic data producers. For all four data sets~\textit{Synthpop} has higher sample fraction equivalence for utility than for risk. \textit{CTGAN} has mixed results with two outcomes of higher risk equivalence than utility (although all equivalences are low compared to the other methods). The effect of different $\epsilon$ values can be observed for~\textit{DataSynthesizer}. For $\epsilon$ = 0.1 the risk and utility equivalence is less than a 0.1\% sample (across all Census datasets); and all but the UK dataset have a negative value for risk (meaning the risk is below the baseline). Considering Figures~\ref{uk1991ru} and \ref{fig2} the~\textit{DataSynthesizer} points have a curvilinear relationship with each other, although where they fall in relation to the sample fractions equivalence varies between the four different Census datasets.

\begin{table}[tb]
    \begin{centering}
    \caption{Synthetic data risk and utility (mean of 5 datasets), and comparable sample equivalence for each of the Census datasets.\protect\footnotemark[6]} \label{tab:results}
    \begin{tabular}{l|lcccc}
        \toprule
        Data    & Synthesizer                   & Overall & Risk            & Sample Equiv. & Sample Equiv. \\
                &                               & Utility & (marginal TCAP) & for Utility   & for Risk \\
        \midrule
                & \textit{CTGAN}                & 0.514   & 0.371       & 0.25\% - 0.5\%    & 2\% - 3\% \\
                & \textit{Synthpop}             & 0.774   & 0.516       & 10\% - 20\%       & 10\% - 20\% \\
                & \textit{DataSynthesizer}:     &         &             &                   &  \\
        UK 1991 & \hspace{2mm}$\epsilon=0.1$    & 0.330   & 0.043       & $<$0.1\%          & $<$0.1\% \\
                & \hspace{2mm}$\epsilon=1$      & 0.416   & 0.303       & $<$0.1\%          & 0.1\% - 0.25\% \\
                & \hspace{2mm}$\epsilon=10$     & 0.536   & 0.424       & 0.25\% - 0.5\%    & 5\% - 10\% \\
                & \hspace{2mm}No DP             & 0.643   & 0.440       & 1\% - 2\%         & 5\% - 10\% \\
        \hline\hline
                & \textit{CTGAN}                & 0.495   & 0.165   & 0.25\% - 0.5\% & $<$0.1\% \\
                & \textit{Synthpop}             & 0.830   & 0.294   & 20\% - 30\%    & 2\% - 3\% \\
        Canada  & \textit{DataSynthesizer}:     &         &         &                &  \\
        2011    & \hspace{2mm}$\epsilon=0.1$    & 0.342   & -0.102  & $<$0.1\%       & $<$0.1\% \\
                & \hspace{2mm}$\epsilon=1$      & 0.425   & 0.011   & 0.1\% - 0.25\% & $<$0.1\% \\
                & \hspace{2mm}$\epsilon=10$     & 0.521   & 0.126   & 0.25\% - 0.5\% & $<$0.1\% \\
                & \hspace{2mm}No DP             & 0.688   & 0.231   & 3\% - 4\%      & 1\% - 2\% \\
        \hline\hline
                & \textit{CTGAN}                & 0.469   & 0.439   & 0.1\% - 0.25\% & 3\% - 4\% \\
                & \textit{Synthpop}             & 0.816   & 0.555   & 20\% - 30\%    & 10\% - 20\% \\
        Fiji    & \textit{DataSynthesizer}:     &         &         &                &  \\
        2007    & \hspace{2mm}$\epsilon=0.1$    & 0.301   & -0.173  & $<$0.1\%       & $<$0.1\% \\
                & \hspace{2mm}$\epsilon=1$      & 0.360   & 0.233   & $<$0.1\%       & $<$0.1\% \\
                & \hspace{2mm}$\epsilon=10$     & 0.477   & 0.414   & 0.1\% - 0.25\% & 2\% - 3\% \\
                & \hspace{2mm}No DP             & 0.727   & 0.526   & 5\% - 10\%     & 5\% - 10\% \\
        \hline\hline
                & \textit{CTGAN}                & 0.430   & 0.412   & 0.5\% - 1\%    & 0.25\% - 0.5\% \\
                & \textit{Synthpop}             & 0.752   & 0.437   & 20\% - 30\%    & 1\% - 2\% \\
        Rwanda  & \textit{DataSynthesizer}:     &         &         &                &  \\
        2012    & \hspace{2mm}$\epsilon=0.1$    & 0.203   & -0.404  & $<$0.1\%       & $<$0.1\% \\
                & \hspace{2mm}$\epsilon=1$      & 0.259   & -0.045  & $<$0.1\%       & $<$0.1\% \\
                & \hspace{2mm}$\epsilon=10$     & 0.373   & 0.230   & 0.1\% - 0.25\% & $<$0.1\% \\
                & \hspace{2mm}No DP             & 0.720   & 0.413   & 10\% - 20\%    & 0.25\% - 0.5\% \\
        \bottomrule
    \end{tabular}
    
    \end{centering}
\end{table}

\section{Discussion}
\label{Discussion}

\footnotetext[6]{Note that the values of $\epsilon$ indicated here are per dataset created, not for the whole synthesis process which will be five times as large given that m=5}

The initial results are very interesting in several respects. 
Firstly the risk-utility relationship for sample data is curvilinear. With risk dropping fast at first as the sample reduces before utility declines rapidly with smaller sample fractions. This is of course ideal and if repeated on larger trial would be a vindication of the use of sampling as a default disclosure control for Census microdata. The curve also indicates a sweet spot sample fraction is around 2-3\%, below this level there is little risk benefit and a large decrease in utility. There is a big caveat to place on this finding, which we will come to shortly.
Second, the results are mixed when comparing synthetic data and samples, with outcomes appearing to vary by country. \textit{Synthpop} generally performs well with the datasets it produced in each country (other than UK) falling to the right of the risk utility curve for the samples. \textit{CTGAN} produced two results on the curve and two to the left of the curve as did \textit{DataSynthesiser}. This would need a more thorough investigation before conclusions could be reached but the driver is presumably variations in data structure.

Third, the impact of varying $\epsilon$ in data synthesiser was also curvilinear but in the opposite less favourable direction (utility decreasing first). Simply switching the differential privacy option on (but with a high value of $\epsilon$) causes a substantial decrease in utility with little appreciable impact on risk. The often advised level of $\epsilon=0.1$ produces datasets that are right down in the left hand corner, with little utility and no risk. This result if validated through larger scale studies would vindicate that impression analysts have about the impact of differential privacy.
The above findings must be strongly caveated on three points.
\begin{enumerate} 
\item The experiments were conducted using samples of microdata. The experimental samples were in fact sub-samples. The results may not generalise to full population data (i.e. we should not assume that sub sample to sample relationships will be replicated in sample to population ones). The true test will be to compare synthetic populations with microdata samples. 
\item The study underestimates the risk of samples relative to synthetic data in general. While we might reasonably assume that synthetic data do not contain identification risk, this is not true for samples (by virtue of them being drawn from real data).  
\item The risk measure employed here uses a response knowledge attribution disclosure. This is sound for Census data but for other datasets, presence detection might be a significant risk that would need to be taken account of. In further work we will be examining this issue further. 
\end{enumerate}

\section{Conclusion}
\label{Conclusion}

This paper has introduced the notion of sample fraction equivalence risk and utility. With experiments using Census data from four countries, we have demonstrated a mechanism for comparison of data synthesis and sampling for microdata. A second subsidiary aim was to bring differential privacy into the same evaluation framework.

The results of the experiments are quite compelling and illustrate the value of the approach. In future work we will be aiming to extend this initial study in three ways: (i)~to run experiments on full population data, (ii)~to assess other disclosure control methods for sample fraction equivalence, and (iii)~to integrate a measure of re-identification disclosure risk into the framework.

\noindent \textbf{Acknowledgements} The authors wish to acknowledge IPUMs International and the statistical offices that provided the underlying data making this research possible: Statistics Canada; Bureau of Statistics, Fiji; National Institute of Statistics, Rwanda; and the Office for National Statistics, UK.

\bibliographystyle{abbrvnat}
\bibliography{references}  

\section*{Appendices}
\subsection*{Appendix A}
\label{AppendA}
A brief summary of the Census microdata:

\textbf{Canada 2011:} Subsetted on the province of Manitoba, containing 32,149 records (3.47\% of the total available dataset, which was a 2.78\% sample of the 2011 Census). Downloaded from IPUMs~\cite{IPUMsCensus}, courtesy of Statistics Canada.

\textbf{Fiji 2007:} The entire 10\% sample (n=84,323) of the 2007 Fiji Census. Downloaded from IPUMs~\cite{IPUMsCensus} courtesy of the Bureau of Statistics, Fiji.

\textbf{Rwanda 2012:} Subsetted on the Karongi region, containing 31,455 records (3.03\% of the total available dataset, which was a 10\% sample of the 2012 Census). Downloaded from IPUMs~\cite{IPUMsCensus} courtesy of the National Institute of Statistics, Rwanda.

\textbf{UK 1991:} Subsetted on the region of West Midlands, containing 104,267 records (9.34\% of total available dataset, which was a 2\% sample of the 1991 Individual Sample of Anonymised Records for the British Census). Downloaded from UK Data Service~\cite{ONSCensusSARs}.

\subsection*{Appendix B}
\label{AppendB}
Summary of TCAP key/target variables. The six key variables are listed together; the first three were used in the case of 3 keys, first four for 4 keys, etc.

\textbf{Canada 2011:}
For target variables (RELIG, CITIZEN and TENURE) the key variables were: AGE, SEX, MARST (marital status), MINORITY (part of a visible minority), EMPSTAT (labour force status), BPL (birthplace).

\textbf{Fiji 2007:}
For target variables (RELIGION, WORKTYPE and TENURE) the key variables were: PROVINCE (of residence), AGE, SEX, MARST (marital status), ETHNIC (part of a visible minority), CLASSWKR (employment status). 

\textbf{Rwanda 2012:}
For target variables (RELIGION, EMPSECTOR and OWNERSH (tenure)) the key variables were: AGE, SEX, MARST (marital status), CLASSWK (employment status), URBAN (urban/rural area), BPL (birthplace).

\textbf{UK 1991:}
For target variables (LTILL (long-term illness), FAMTYPE and TENURE) the key variables were: AREAP, AGE, SEX, MSTATUS (marital status), ETHGROUP (ethnic group), ECONPRIM (economic status).

\subsection*{Appendix C}
\label{AppendC}
Description of regression models used to calculate the CIO. For each dataset two logistic regressions were performed using marital status and housing tenure as the targets (a binary target was created). Eight predictors were used, these were the same for both models (with tenure/marital status removed accordingly):

\textbf{Canada} predictors: ABIDENT (aboriginal identity), AGE, CLASSWK, DEGREE, EMPSTAT, SEX, URBAN, TENURE/MARST

\textbf{Fiji} predictors: AGE, CLASSWKR, ETHNIC, RELIGION, EDATTAIN (educational level attained), SEX, PROVINCE, TENURE/MARST

\textbf{Rwanda} predictors: AGE, DISAB1, EDCERT (highest educational qualification), CLASSWK, LIT (languages spoken), RELIG, SEX, TENURE/MARST

\textbf{UK} predictors: AGE, ECONPRIM, ETHGROUP, LTILL, QUALNUM, SEX, SOCLASS, TENURE/MSTATUS

\renewcommand{\thefigure}{D\arabic{figure}}
\setcounter{figure}{0}

\newpage
\subsection*{Appendix D}
\label{AppendD}

\begin{figure}[hb]
\includegraphics[width=\textwidth]{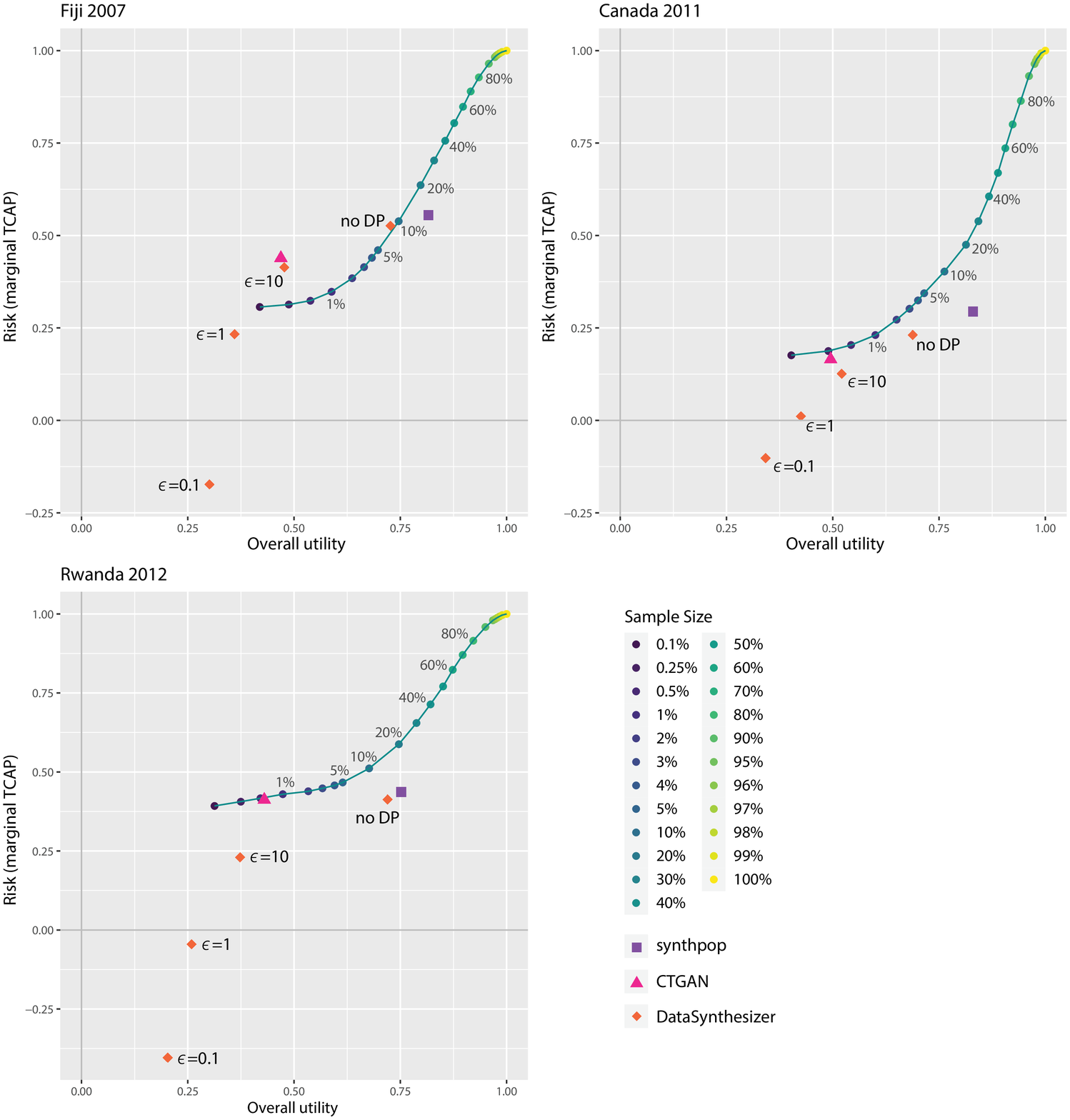}
\caption{Risk-Utility map plotting the mean synthetic data and sample fraction results for Fiji 2007, Canada 2011 and Rwanda 2012 Census data.} \label{fig2}
\end{figure}

\end{document}